\pdfoutput=1

\documentclass[11pt]{article}

\usepackage[preprint]{acl}
\usepackage{multirow}
\usepackage{times}
\usepackage{latexsym}
\usepackage{booktabs}   
\usepackage{amssymb}    

\newtheorem{definition}{Definition}[section] 
\usepackage[T1]{fontenc}

\usepackage[utf8]{inputenc}

\usepackage{microtype}
\usepackage{amsmath}
\usepackage{inconsolata}
\usepackage{subfig}
\usepackage{subfig}
\usepackage{graphicx}
\usepackage{tcolorbox}
\tcbuselibrary{skins} 
\usepackage{lipsum} 
\usepackage{xspace}
\usepackage{enumitem}
\newcommand{\ourmethod}{\textsc{{EIB-Learner}}\xspace}

\definecolor{deepblue}{RGB}{0, 0, 139}
\definecolor{deepgreen}{RGB}{0, 100, 0}   
\definecolor{deepred}{RGB}{139, 0, 0}     
\definecolor{deeppurple}{RGB}{75, 0, 130}  

\usepackage{booktabs} 

\usepackage{amsmath,amssymb,amsfonts}

\usepackage[ruled,vlined,noend]{algorithm2e}

\SetAlgoNoLine

\SetCommentSty{CommentFont}

\SetKwComment{Comment}{/* }{ */}

\title{Understanding the Information Propagation Effects of Communication Topologies in LLM-based Multi-Agent Systems}

\author{
\textbf{Xu Shen\textsuperscript{1}\footnotemark[1]},
 \textbf{Yixin Liu\textsuperscript{2}\footnotemark[1]},
 \textbf{Yiwei Dai\textsuperscript{1}},
 \textbf{Yili Wang\textsuperscript{1}},
 \textbf{Rui Miao\textsuperscript{1}},
 \textbf{Yue Tan\textsuperscript{3}},
 \textbf{Shirui Pan\textsuperscript{2}},
 \textbf{Xin Wang\textsuperscript{1}\footnotemark[2]},
 \\
 \textsuperscript{1}School of Artificial Intelligence, Jilin University, Changchun, China,
 \\
 \textsuperscript{2}School of Information and Communication Technology, Griffith University, Goldcoast, Australia,
 \\
 \textsuperscript{3}School of Computer Science and Engineering, University of New South Wales, Sydney, Australia
 \\
 \{\texttt{shenxu23}, \texttt{daiyw23}, \texttt{wangyl21}, \texttt{ruimiao20}\}@mails.jlu.edu.cn, 
 \{\texttt{yixin.liu}, \texttt{s.pan}\}@griffith.edu.au, \\
 \{\texttt{yue.tan}\}@unsw.edu.au, 
 \{\texttt{xinwang}\}@jlu.edu.cn
}

\begin{document}
\maketitle
\footnotetext[1]{Equal Contribution}
\footnotetext[2]{Corresponding Author}
\begin{abstract}
The communication topology in large language model-based multi-agent systems fundamentally governs inter-agent collaboration patterns, critically shaping both the efficiency and effectiveness of collective decision-making. While recent studies for communication topology automated design tend to construct sparse structures for efficiency, they often overlook why and when sparse and dense topologies help or hinder collaboration. 
In this paper, we present a causal framework to analyze how agent outputs, whether correct or erroneous, propagate under topologies with varying sparsity. Our empirical studies reveal that moderately sparse topologies, which effectively suppress error propagation while preserving beneficial information diffusion, typically achieve optimal task performance. 
Guided by this insight, we propose a novel topology design approach, \ourmethod, that balances error suppression and beneficial information propagation by fusing connectivity patterns from both dense and sparse graphs. Extensive experiments show the superior effectiveness, communication cost, and robustness of \ourmethod. The code is in: \href{https://github.com/se7esx/EIB/}{https://github.com/se7esx/EIB/}.
\end{abstract}

\section{Introduction}

Agents powered by large language models (LLMs) ~\cite{li2023camel,wang2024survey,xi2025rise} have demonstrated strong performance across tasks like reasoning~\cite{yao2023react}, code generation~\cite{zhang2024codeagent}, and complex decision-making~\cite{guo2024embodied}. A key advancement in this area is the development of collaborative \textbf{multi-agent systems (MAS)}, where multiple LLM agents interact to outperform single agent when tackling complex tasks~\cite{talebirad2023multi,liang2024encouraging,guo2024large}. While agent role specialization contributes to this improvement~\cite{bo2024reflective}, the communication topology, which regulates how agents \textit{propagate}, \textit{exchange}, and \textit{process} information, plays a more fundamental role in enabling effective multi-agent collaboration~\cite{qian2024scaling,zhang2024exploring}. From simple chain~\cite{zhang2024chain} and tree~\cite{zhou2024language}, to fully connected graphs~\cite{hu2024learning} and LLM-generated designs~\cite{zhuge2024gptswarm}, prior works have proposed various topologies to shape this interaction. These studies converge on a central insight: \textbf{\textit{the topology of communication is critical to the effectiveness and efficiency of MAS}}.
\begin{figure}[t!]   
\vspace{-5mm}
\centering    
	\includegraphics[width=\columnwidth]{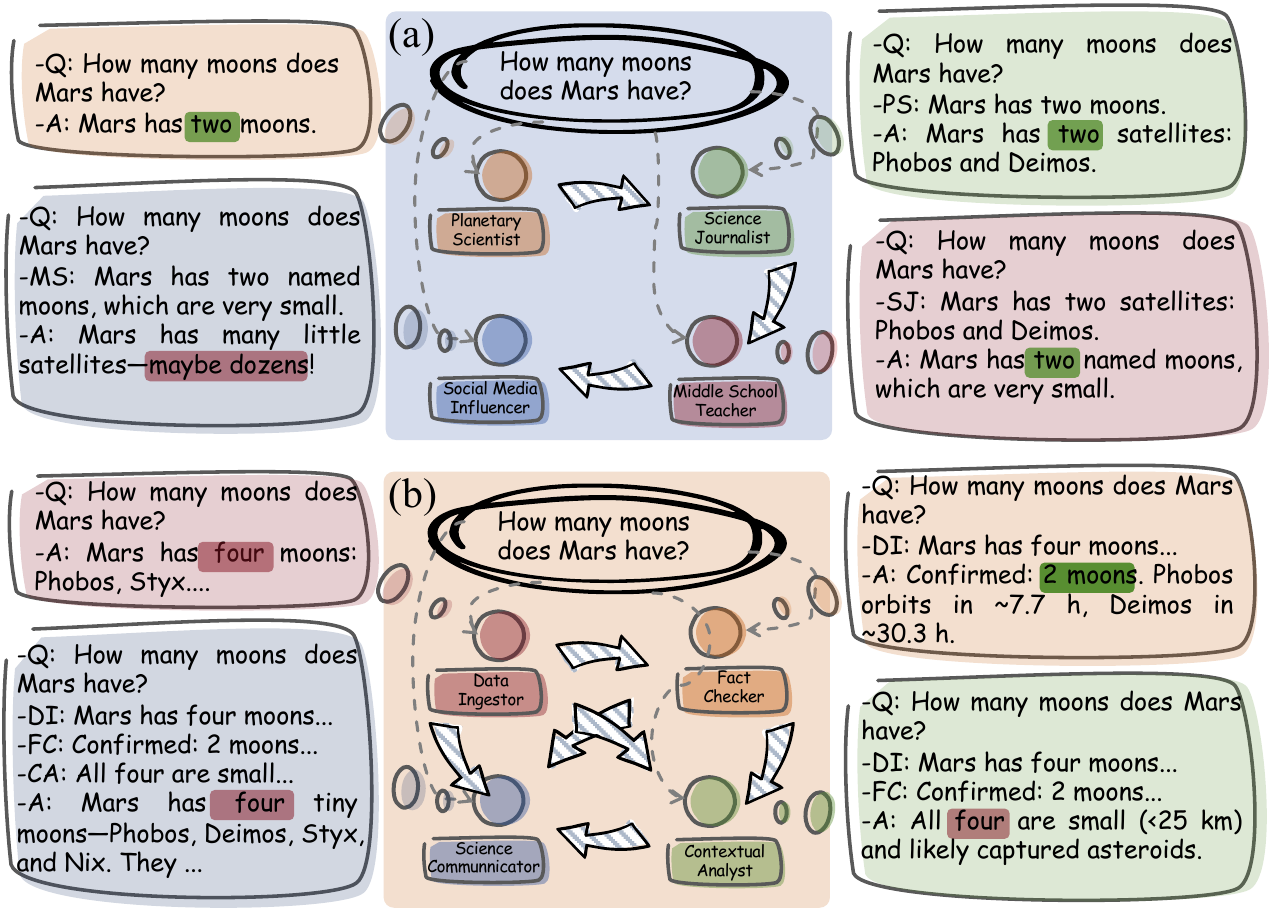}
			\vspace{-6mm}
					\caption{Illustration of (a) insight suppression caused by sparse chain and (b) error propagation induced by dense fully connected topologies.}
					\label{fig:intro}  
\vspace{-5mm}
\end{figure}

Given its critical role in MAS performance, recent studies have increasingly focused on the automated design of communication topology using graph learning techniques~\cite{li2024improving,liu2025advances,zhou2025multi}, advancing beyond using pre-defined topologies. Central to these approaches is the principle of \textbf{\textit{sparsifying}} communication graphs, motivated by the insight that fewer but higher-quality interactions enhance system efficiency. 
For example, AgentPrune~\cite{zhang2024cut} and AgentDrop~\cite{wang2025agentdropout} learn a task-specific sparse graph to eliminate suboptimal connections and/or agents, while G-Designer~\cite{zhang2024g} introduces a graph sparsification learning objective for the communication design model. Despite their advanced performance and efficiency, these methods often treat sparsification as an end goal, without considering a deeper question: \textit{{why} and {when} do \textbf{sparse} topologies help MAS collaboration?} Conversely, despite enabling maximal inter-agent interaction, \textit{why and when do \textbf{dense} topologies underperform}?


To answer the above questions, from a causal perspective, we conduct comprehensive analyses to investigate how the communication topologies with varying sparsity affect the information propagation among agents and thereby impact decision-making performance. 
Through empirical analysis, we find that sparser topologies exhibit greater robustness against the propagation of error information generated by an individual agent, but may also suppress beneficial insights that contribute to accurate collective decision-making (see Figure~\ref{fig:intro}a). Conversely, denser topologies enable thorough insight dissemination, but also aggressively propagate individual errors (see Figure~\ref{fig:intro}b). With task-oriented analysis, we derive a novel insight: \textbf{\textit{an ideal communication topology should have a moderate sparsity, effectively suppressing error propagation while maintaining beneficial insight propagation.}}


Based on the insight above, we propose \textbf{E}rror-\textbf{I}nsight \textbf{B}alanced \textbf{Learner} (\ourmethod for short) for automated communication topology design of LLM-based MAS. Given a specific query, \ourmethod can dynamically customize an optimal topology by balancing error suppression and insight propagation. 
Specifically, \ourmethod utilizes dual-view graph neural networks (GNNs) to simultaneously simulate error suppression on sparse graphs and insight propagation on dense graphs, creating complementary topological representations that inform optimal connectivity. Then, \ourmethod integrates the connectivity coefficients learned by both views with query-aware adaptive fusion, blending the error robustness of sparse topologies with the insight propagation capacity of dense topologies. Extensive experiments on 6 benchmark datasets demonstrate the effectiveness, communication efficiency, and robustness of \ourmethod.
 To sum up, the main contributions of this paper are as follows:
 \begin{itemize}[leftmargin=10pt]
    \item \textbf{Causal Analysis.} We conduct causal counterfactual analyses to assess the impact of an agent on collective decision making in communication topologies with varying sparsity, revealing how error and insight propagation affect the inter-agent collaboration.
    \item \textbf{Novel Method.} We propose \ourmethod, a GNN-based communication topology design approach that simulates error and insight propagation in MAS and learn reliable and efficient topologies by balancing error-insight trade-off.
    \item \textbf{Comprehensive Experiments.} Experiments on 6 MAS decision-making benchmarks demonstrate that \ourmethod achieves superior accuracy with reduced communication cost and better robustness compared to existing baselines.
\end{itemize}
\section{Problem Definition}
\paragraph{{Communication Topology}}
We formulate the communication topology of a multi-agent system (MAS) as a directed acyclic graph (DAG), defined as $\mathcal{G} = (\mathcal{V},\mathcal{E})$. Here $\mathcal{V} = \{v_{1},\cdots,{v}_{N}\}$ denotes the set of nodes, where each node $v_i$ indicates a LLM-based agent. The edge set $\mathcal{E}$ specifies directed communication links, where each edge $ e_{ij} = (v_{i}, v_{j}) \in \mathcal{E}$ indicates that agent $v_{i}$ can receive messages from agent $v_{j}$. The connectivity of $\mathcal{g}$ can be represented by an adjacency matrix $\mathbf{A} \in \{0,1\}^{N \times N}$, where $\mathbf{A}_{i,j} = 1$ if and only if $ (v_{i}, v_{j}) \in \mathcal{E}$ and $\mathbf{A}_{i,j} = 0$ otherwise.

In this setup, each node $v_{i} \in V$ corresponds to a LLM-based agent and can be formalized as
$v_{i} = \{\texttt{Role}_{i},\texttt{State}_{i}\}$, where $\texttt{Role}_{i}$ denotes the pre-defined role of agent $v_{i}$ for a specific task, and $\texttt{State}_{i}$ stores the history response of the agent $v_{i}$ and its interaction history with other agents.  Each LLM-based agent receives a prompt $\mathcal{P}$ and generates the response $\mathcal{R}_{i}$ by:
\begin{equation}
    \mathcal{R}_{i} = v_{i}(\mathcal{P}) = v_{i}(\mathcal{P}_{\text{sys}},\mathcal{P}_{\text{usr}}),
\end{equation}
where $\mathcal{P}_{\text{sys}}$ is the system prompt that combine the $\texttt{Role}_{i}$ and $\texttt{State}_{i}$ of the current agent, and $\mathcal{P}_{\text{usr}}$ denotes the user prompt, including current query, task instructions, and other necessary information.

\paragraph{{Communication Protocol}}
Given a user query $\mathcal{Q}$, a MAS performs $K$ rounds of interaction based on the communication graph $\mathcal{G}$. At each round $t$, the execution order of agents is determined by a topological sort $\sigma = [v_{\sigma_1}, \dots, v_{\sigma_N}]$, ensuring that an agent is only activated after all its in-neighbors have produced their outputs. Each agent $v_{i}$ generates its own response $\mathcal{R}_{i}^{(t)}$ according to the user query and the outputs of its neighbors:
\begin{align} \label{mp_in_multi-agent}
    \mathcal{R}_{i}^{(t)} &= v_{i}(\mathcal{P}_{\text{sys}},\mathcal{P}^{(t)}_{\text{usr}}), \\
    \mathcal{P}^{(t)}_{\text{usr}}&=\{\mathcal{Q}\} \cup \{\mathcal{R}_{j}^{(t)} | v_j\in \mathcal{N}(v_{i})\}, \nonumber
\end{align}
where $\mathcal{N}(v_{i})=\{v_j|(v_j,v_i)\in \mathcal{E})\}$ denotes the neighbor set of the agent $v_{i}$. After $K$ rounds, the final output $\mathcal{O}$ is obtained by aggregating all agent outputs:
\begin{equation}
    \mathcal{O} = \operatorname{Aggregate}(\mathcal{R}_{i}^{(K)} \dots \mathcal{R}_{n}^{(K)}),
\end{equation}
\noindent where the aggregation strategy $\operatorname{Aggregate}(\cdot)$ varies across implementations, including majority voting among agents, delegating the final decision to a specific agent, or selecting the output from the last agent in the execution order. The communication rounds $K$ can be either predefined or adaptively determined via early-stopping mechanisms.

\paragraph{{MAS Communication Topology Design Problem}} 
In this paper, we aim to design a communication topology for a LLM-based MAS that maximizes the utility of the collective output under a given task query. Formally, Our objective is to find an optimal topology $\mathcal{G}^{}$ from a feasible topology space $\mathbb{G}$ such that:
\begin{equation}
\mathcal{G}^{} = \arg\max_{\mathcal{G} \in \mathbb{G}} \phi(\mathcal{G}(\mathcal{Q})),
\end{equation}
where $\phi(\cdot)$ denotes a utility function that evaluates the correctness or overall quality of the system output $\mathcal{G}(\mathcal{Q})$ in response to the input query $\mathcal{Q}$. The search space $\mathbb{G}$ includes all valid DAGs over the agent set $\mathcal{V}$ that satisfy acyclicity and predefined execution constraints.

\section{Analysis of Communication Topology}

In this section, we conduct extensive empirical analysis to investigate how communication topologies of varying sparsity shape the actual information propagation among agents and influence collective decision-making outcomes in a multi-agent system (MAS). \textit{First}, we discuss two causal effects of communication topologies with different sparsity: \textbf{\textit{error propagation}} and \textbf{\textit{insight propagation}}. Specifically, we analyze how the erroneous information generated by an individual agent spreads, as well as how beneficial insights diffuse across the topology. \textit{Next}, we evaluate how sparsity influences the effectiveness through task-oriented performance analysis, and expose the ideal information propagation characteristics of communication topologies. \textit{Finally}, we assess the information diffusion capabilities of existing communication paradigms, including both pre-defined and automatically learned topologies, with respect to their error and insight propagation properties.

\subsection{Causal Metric of Agent Impact}


In order to analyze the information propagation effects of different communication topologies, we first define a new causal metric, denoted as Counterfactual Agent Propagation Effect (\textbf{CAPE}), that quantifies the causal influence of the output of a single agent on the final answer of MAS under a given communication graph. 
Specifically, we perform a counterfactual intervention by forcing agent $v_{i}$ to produce a deliberately manipulated output, and then measure whether this change alters the final prediction made by the MAS under a fixed topology $\mathcal{G}$. The formal definition is as follows:
\begin{definition}
   Given a MAS under communication graph $\mathcal{G}$, for a query $\mathcal{Q}$, let $y_{Q}^\text{orig} \in \{0,1\}$ denote whether the final system answer is correct under the original communication process. The Counterfactual Agent Propagation Effect {\text{(CAPE)}} of agent $v_i$ under $\mathcal{G}$ and query $\mathcal{Q}$ as:
   \begin{equation}
       \resizebox{.8\hsize}{!}{$\text{CAPE}_{i}(\mathcal{G},\mathcal{Q}) = \mathbb{I}[y_q^{\text{cf}}(\mathcal{G}, \text{do}(\mathcal{O}_i := \hat{\mathcal{{O}}}_i)) \ne y_q^{\text{orig}}(\mathcal{G})]$},
   \end{equation}
   where $ \operatorname{do}(\mathcal{O}_i := \hat{\mathcal{{O}}}_i))$ denotes a targeted intervention that forces only agent $v_{i}$ output to a counterfactual value $\hat{\mathcal{{O}}}_i$, $y_{\mathcal{Q}}^{\text{cf}} \in \{0,1\}$ is the correctness of the final system prediction after the intervention, and $\mathbb{I} \left[ \cdot
\right]$ is the indicator function for whether the final answer of MAS is flipped.
\end{definition}

A higher {CAPE} value suggests that changes in the output of an individual agent, once propagated through the communication structure, have a stronger influence on the final decision of MAS. Building on the ability of CAPE to quantify agent-level influence, we further define the Total Counterfactual Topology Effect (\textbf{TCTE}) to expend the casual metric to the topology level, providing a comprehensive measure of the general sensitivity of the topology to localized interventions.
\begin{definition}
Given a communication topology $\mathcal{G}$, its Total Counterfactual Topology Effect ({TCTE}) given a query $\mathcal{Q}$ is defined as:
\begin{equation}
\text{TCTE}(\mathcal{G},\mathcal{Q}) = \frac{1}{N}\sum_{v_i\in \mathcal{V}} \frac{1}{\sqrt{d_i}} \cdot \text{CAPE}_{i}(\mathcal{G},\mathcal{Q}),
\end{equation}
where $d_i$ is the degree of agent $v_i$ in graph $\mathcal{G}$. 
\end{definition}

In TCTE, the coefficient $1/\sqrt{d_i}$ downweights the agents with high degree, reflecting the intuition that highly connected agents are more likely to spread influence, and we wish to normalize their impact accordingly. A higher TCTE indicates that the topology is more responsive to specific changes in agent-level output, implying a greater potential to either propagate erroneous information or effectively transmit correct insights.

\subsection{Empirical Analysis}
In this subsection, we aim to empirically assess how different communication topologies affect the sensitivity of MAS to agent-level perturbations. We quantify the sensitivity by computing TCTE of communication graphs, and also evaluate the effectiveness of different topologies using task-oriented measurement, i.e., question-answering accuracy.

\begin{figure}[t]
  \centering
  \subfloat[Error propagation]{%
    \label{subfig:error}%
    \includegraphics[
      width=0.5\columnwidth,
    ]{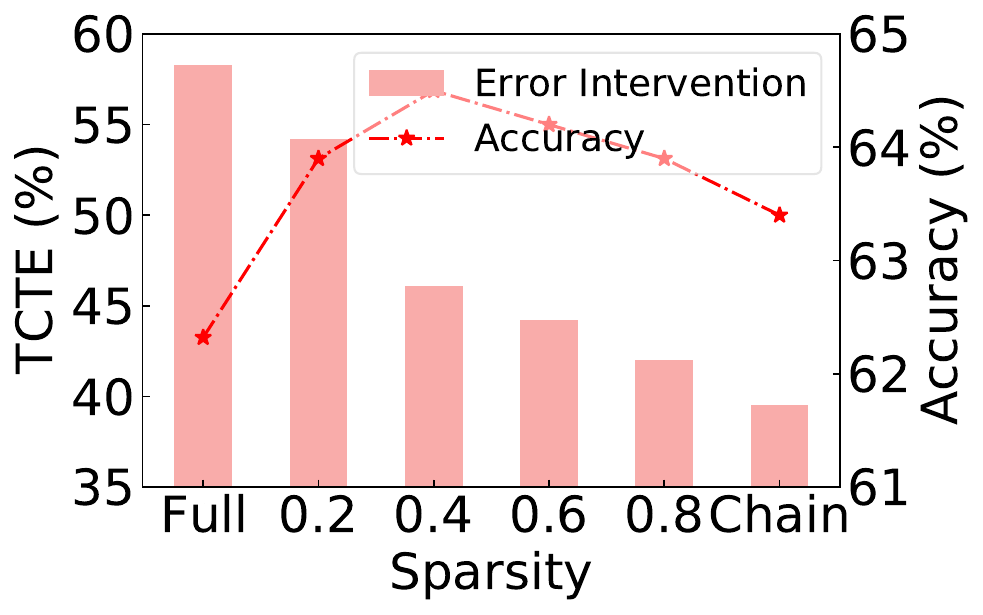}%
  }\hfill
  \subfloat[Insight propagation]{%
    \label{subfig:insight}%
    \includegraphics[
      width=0.5\columnwidth,
    ]{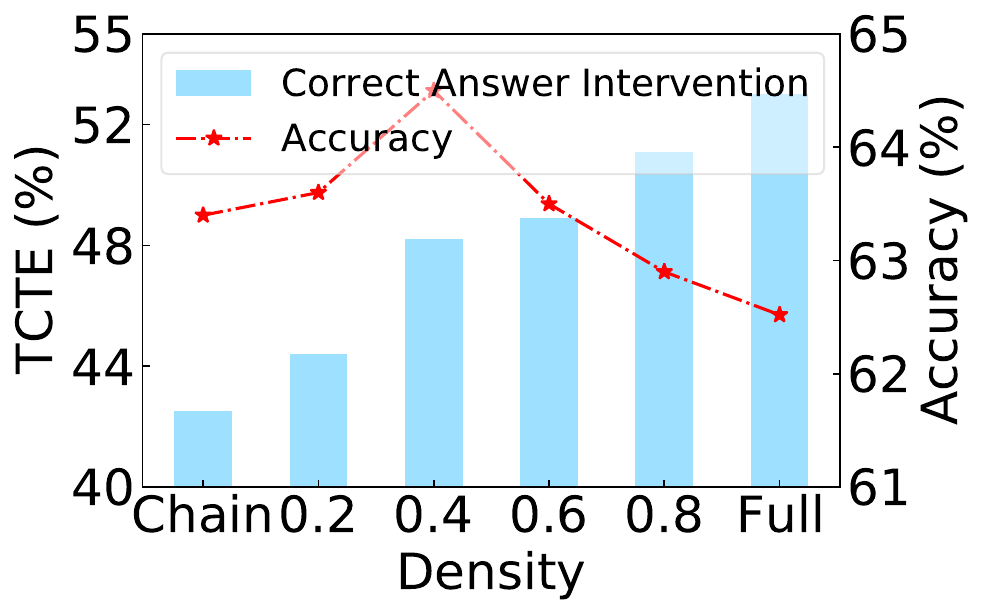}%
  }

  \vspace{-1em} 

  \subfloat[Analysis of pre-defined and learned topologies]{%
    \label{subfig:baselinetp}%
    \includegraphics[
      width=\columnwidth
    ]{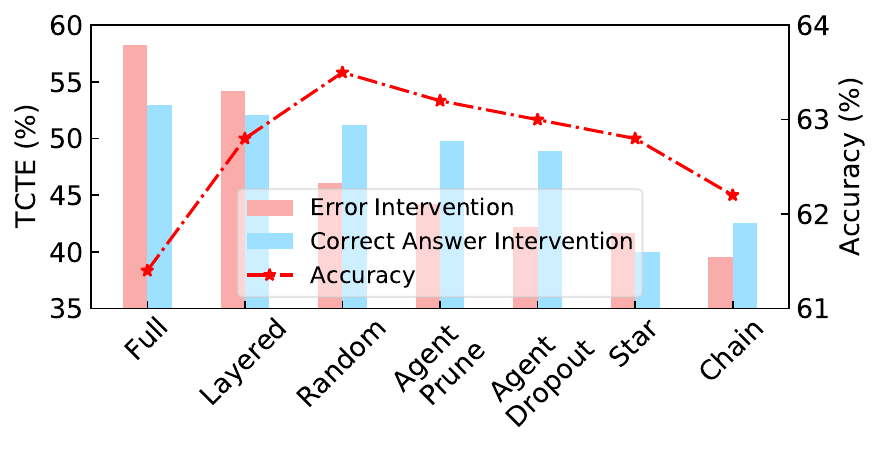}%
  }

  \caption{Empirical results of error propagation effect and insight propagation effects.}
  \label{fig:three_plots}
  \vskip -1 em
\end{figure}

\paragraph{Experimental Settings} 
We conduct our experiments on the MMLU dataset~\cite{hendrycks2020measuring}, a benchmark for evaluating knowledge and reasoning across diverse domains. In our MAS setup, each of the six agents is an independent instance of GPT-3.5 with a distinct role and prompt template, and all communicate according to a pre-specified topology. For each topology, we construct two evaluation sets: one containing 500 questions that the MAS originally answers correctly, and another with 500 questions that the system answers incorrectly. These sets serve as the basis for simulating both detrimental and beneficial agent-level interventions under varying graph structures.

\subsubsection{Analysis of Error Propagation Effect}

\paragraph{Setup} 
To investigate \textbf{\textit{how different communication topologies spread the erroneous information generated by a single agent}}, we focus on the 500 originally correct questions and assess the vulnerability of MAS to local errors. Specifically, we start from a fully connected communication graph (\texttt{Full}) and progressively remove edges randomly to generate topologies with different sparsity levels, until it degrades into a chain graph (\texttt{Chain}). For each graph, we apply a counterfactual intervention by modifying the system prompt of an agent to produce an incorrect output, i.e., $\hat{\mathcal{{O}}}_i = \mathcal{O}_{\text{error}}$, and compute the corresponding $\text{TCTE}$ to quantify how often the final prediction flips from correct to incorrect under each topology. 

\paragraph{Discussion of Results} 
As shown in Figure \ref{subfig:error}, dense topologies exhibit greater susceptibility to error propagation, wherein erroneous outputs from individual agents are more likely to compromise the final prediction. When the sparsity is zero (i.e., \texttt{Full}), $\text{TCTE}$ reaches its peak. In contrast, sparser structures can effectively mitigate this issue. For example, in the extreme case \texttt{Chain}, $\text{TCTE}$ is reduced by up to $19\%$, indicating improved robustness to localized errors. 
\begin{tcolorbox}[
    enhanced, 
    colback=white, 
    colframe=deepgreen, 
    boxrule=1pt, 
    arc=5pt, 
    left=5pt, right=5pt, 
    top=5pt, bottom=5pt, 
    width=\linewidth 
]
\textbf{\textcolor{deepgreen}{Finding 1}}: Denser communication topologies are more susceptible to error spreading, while sparser topologies demonstrate stronger resistance to the error propagation effect. 
\end{tcolorbox}

\subsubsection{Analysis of Insight Propagation Effect}

\paragraph{Setup} 
In this experiment, we examine \textit{\textbf{how different topologies support the propagation of beneficial insight generated by each agent}}. We focus on the 500 originally incorrect questions and start from a sparse \texttt{Chain} topology, gradually adding edges to increase connectivity until reaching \texttt{Full}. In each case, we perform a counterfactual intervention by injecting the correct answer into a selected agent’s prompt, i.e., $\hat{\mathcal{{O}}}_i = \mathcal{{O}}_{\text{correct}}$, and compute the corresponding $\text{TCTE}$ to measure how often the final prediction flips from incorrect to correct.

\paragraph{Discussion of Results} 
As shown in Figure \ref{subfig:insight}, sparse topologies tend to impede the propagation of accurate and informative signals, preventing them from influencing the final output. When connectivity is minimal (i.e., \texttt{Chain}), $\text{TCTE}$ reaches its lowest value, indicating poor transmission of correct insights. With increasing density, the insight propagation effect is strengthened. In \texttt{Full} structure, $\text{TCTE}$ increases by $10.5\%$, suggesting that the system becomes more capable of integrating accurate agent-level inputs into the final decision. 

\begin{tcolorbox}[
    enhanced, 
    colback=white, 
    colframe=deepgreen, 
    boxrule=1pt, 
    arc=5pt, 
    left=5pt, right=5pt, 
    top=5pt, bottom=5pt, 
    width=\linewidth 
]
\textbf{\textcolor{deepgreen}{Finding 2}}: Increased connectivity enhances beneficial insight propagation in MAS, while sparse topologies constrain the integration of informative signals from individual agents.
\end{tcolorbox}


\subsubsection{Effectiveness Analysis}

\paragraph{Setup} 
While communication topologies of different sparsity levels produce distinct error propagation and insight propagation effects, a follow-up question raises: \textbf{\textit{How do the topological sparsity ultimately affect real-world task performance?}} To answer this question, we examine the task-oriented performance of different topologies, in term of accuracy for MMLU dataset.

\paragraph{Discussion of Results} 
The task-oriented accuracies are demonstrated by the red dashed line in Figures~\ref{subfig:error} and \ref{subfig:insight}. From both cases, we can observe that a moderate sparsity contributes to better task performance, while overly sparse or dense structures lead to suboptimal outcomes. Recalling \textbf{\textcolor{deepgreen}{Finding 1}} and \textbf{\textcolor{deepgreen}{Finding 2}}, this phenomenon arises because topology with intermediate sparsity strikes a balance: it sufficiently suppresses erroneous information while effectively propagating beneficial insights, thereby maximizing the decision-making accuracy of MAS. At the agent level, an ideal topology should balance this trade-off through moderate sparsity: establishing dense connections around high-accuracy agents to amplify beneficial insights, while maintaining sparse connectivity for error-prone agents to limit misinformation propagation.

\begin{tcolorbox}[
    enhanced, 
    colback=white, 
    colframe=deepblue, 
    boxrule=1pt, 
    arc=5pt, 
    left=5pt, right=5pt, 
    top=5pt, bottom=5pt, 
    width=\linewidth 
]
\textbf{\textcolor{deepblue}{Insight}}: Topologies with moderate sparsity optimize MAS performance by balancing error suppression and insight propagation, achieved through dense connections around accurate agents and sparse links for error-prone agents.
\end{tcolorbox}

\subsubsection{Analysis of Existing Topologies}
\paragraph{Setup} 
With the \textbf{\textcolor{deepblue}{Insight}} given above, we are curious about: \textbf{\textit{How do existing communication topologies, both pre-defined and learnable, perform in balancing error propagation and insight diffusion?}} 
We evaluate a range of representative topologies, including pre-defined ones such as \texttt{Full}, \texttt{Layered}, \texttt{Random}, \texttt{Star}, and \texttt{Chain}, as well as topologies learned through state-of-the-art methods such as \texttt{AgentDropout}~\cite{wang2025agentdropout} and \texttt{AgentPrune}~\cite{zhang2024cut}. 
\vspace{-6pt}
\paragraph{Discussion of Results} 
As shown in Figure \ref{subfig:baselinetp}, the results reveal a consistent trend: sparse topologies are more robust to error propagation, while denser structures better support the amplification of helpful insight, which echoes our earlier analysis. However, we found that the topologies generated by \texttt{AgentDrop} and \texttt{AgentPrune} do not show superior performance compared to random graphs with moderate sparsity, indicating that topology optimization methods that explicitly balance error and insight propagation. 

\begin{tcolorbox}[
    enhanced, 
    colback=white, 
    colframe=purple, 
    boxrule=1pt, 
    arc=5pt, 
    left=5pt, right=5pt, 
    top=5pt, bottom=5pt, 
    width=\linewidth 
]
\textcolor{purple}{\textbf{Limitation}}: Current methods show limited effectiveness in simultaneously managing error suppression and insight propagation. 
\end{tcolorbox}

 \begin{figure*}[ht!]    
					\centering    
	\label{fig:subfig5}\includegraphics[width=1.8\columnwidth]{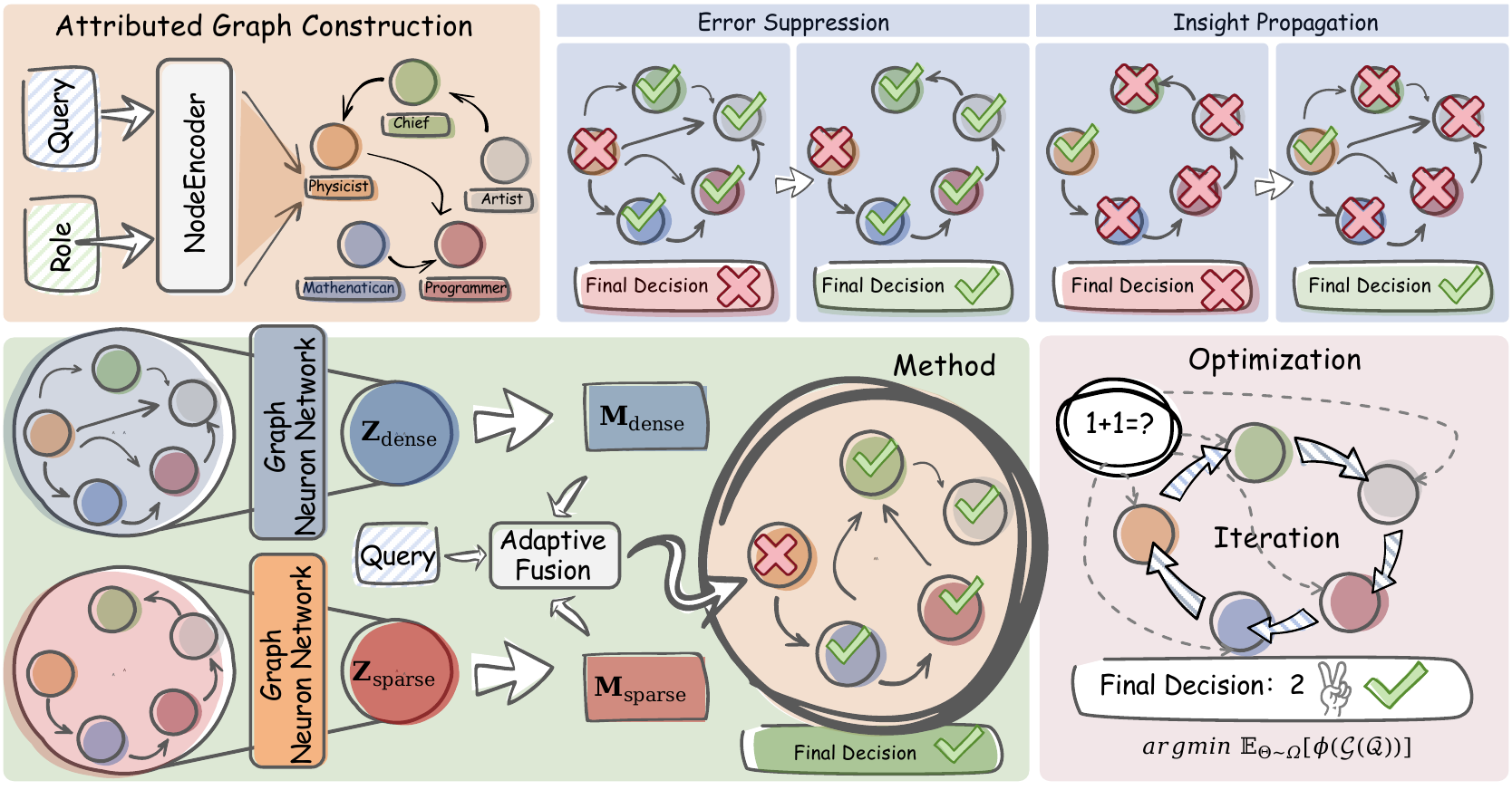}
			\vspace{-4mm}
					\caption{The overall framework of \ourmethod. \ourmethod simulates MAS communication via GNNs, generating two connectivity coefficient matrices to suppress error spreading (sparse view) and enhance insight propagation (dense view), which are then combined by a query-aware fusion module into an optimal topology.}
                    
					\label{fig:framework}            
                    \vspace{-3mm}
				\end{figure*}

\section{Methodology}
To address the above \textcolor{purple}{\textbf{Limitation}}, in this section, we propose a novel
\textbf{E}rror-\textbf{I}nsight \textbf{B}alanced \textbf{Learner} (\ourmethod for short) for communication topology optimization in multi-agent system (MAS). Motivated by our \textcolor{deepblue}{\textbf{Insight}}, \ourmethod balances the error-insight trade-off by co-training two complementary graph neural network (GNN) simulators to simulate the error suppression and insight propagation given a specific query (Section~\ref{GNN}), and then adaptively blending their learned inter-agent coefficients to construct robust topologies (Section~\ref{dual_graph}). The overall pipeline of \ourmethod is shown in Figure~\ref{fig:framework}.


\subsection{GNN-based Propagation Simulators}\label{GNN}

To balance error suppression and insight propagation in communication topologies, a critical prerequisite is identifying beneficial agents (those likely to contribute accurate insights) and error-prone agents (those susceptible to generating misinformation) within the MAS for a given query $\mathcal{Q}$, and then model the information flow among these agents. Nevertheless, without prior knowledge of agent reliability in answering $\mathcal{Q}$, it is challenging to distinguish the agents and simulate their responses. To address this challenge, we utilize GNNs to simulate the responding process of MAS given a query, and hence model error and insight propagation.

\paragraph{MAS as an Attributed Graph}
While the inter-agent communication can be represented as a graph structure, we further embed the agent role information and query text into the feature input of GNNs. Specifically, the feature of each agent node $v_i$ can be acquired by:
\begin{equation}
	\mathbf{x}_i \leftarrow \text{NodeEncoder}(\mathcal{T}(\texttt{Role}_{i}), \mathcal{Q}),
\end{equation}
where $\mathcal{T}(\cdot)$ extracts the textual description of the agent role and its corresponding prompt, and $\text{NodeEncoder}(\cdot)$ can be implemented using lightweight pre-trained text embedding models. The integration of agent role and query allows the GNN model to capture the contribution (beneficial or error-prone) of each agent given a specific query.

\vspace{-7pt}
\paragraph{GNNs for Propagation Simulation}
With the MAS attributed graph as input, we leverage GNNs to simulate the information propagation in MAS. 
Formally, at the $l$-th GNN layer, the message representation for node $v_i$ is updated by aggregating features from its neighbors:
\vspace{-5pt}
\begin{equation}\label{message-passing}
\resizebox{.85\hsize}{!}{$\mathbf{m}_{i}^{(l)} = \operatorname{MP}\left (\mathbf{m}_{i}^{(l-1)},\{ \mathbf{m}_{j}^{(L-1)} \mid v_j \in \mathcal{N}(v_i) \} \right),$}
\vspace{-7pt}
\end{equation}
where $\mathbf{m}^{(0)}_i = \mathbf{X}_i$ is the raw input feature of node $v_i$, $\mathcal{N}(v_i)$ denotes the set of neighbors, and $\operatorname{MP}(\cdot)$ is the message passing function that aggregates and transforms neighboring information. The message passing scheme of GNNs can effectively mimic the information flow in MAS: the message aggregation of neighbors reflects the communication between agents, while the feature transformation can simulate the agent responding given a query. 

In \ourmethod, we design a dual-view framework, with two GNNs to simulate the error suppression and insight propagation, respectively. Inspired by \textbf{\textcolor{deepgreen}{Finding 1}}, we employ the most sparse \texttt{chain} topology to simulate error suppression in MAS; In the other hand, \textbf{\textcolor{deepgreen}{Finding 2}} motivates us to use the densest \texttt{full} topology to simulate insight propagation effect. These two different topologies determine different neighbor definitions in Eq.~\eqref{message-passing}.



\subsection{Error-Insight Balanced Topology Learner}\label{dual_graph}


\paragraph{Inter-Agent Coefficient Modeling}
Building upon the GNN simulators, we use an inner-product decoder to estimate pairwise connectivity to model the inter-agent connectivity in the optimal communication topology, which can be written as:
\begin{equation}
\resizebox{.85\hsize}{!}{$
    \mathbf{Z}_{\text{view}} = \operatorname{GNN}(\mathbf{A}_{\text{view}},\mathbf{X}), \mathbf{M}_{\text{view}} =  \sigma(\mathbf{Z}_{\text{view}}^{T}\mathbf{Z}_{\text{view}}),$}
\end{equation}
where subscript ``view'' can be ``sparse'' or ``dense'', corresponding to the views that simulate error suppression and insight propagation, respectively, $\mathbf{M}_{\text{view}}$ is the coefficient matrix to model inter-agent connectivity, and $\sigma(\cdot)$ denotes the sigmoid function to ensure outputs are in $[0, 1]$. The inner product operation leverages propagated information to estimate agent compatibility, determining the optimal likelihood of connection between any two agents in the communication topology. 


\paragraph{Adaptive Dual-View Fusion}
Based on our \textcolor{deepblue}{\textbf{Insight}}, the optimal communication topology should balance the factors of error suppression and insight propagation, which are modeled by the sparse and dense views, respectively. To this end, we introduce a simple yet effective fusion function to fuse $\mathbf{M}_{\text{sparse}}$ and $\mathbf{M}_{\text{dense}}$ into a comprehensive coefficient matrix $\mathbf{M}_{\text{final}}$.

Concretely, we implement the fusion with a query-aware gating mechanism. Firstly, a lightweight feedforward network (i.e. $\operatorname{MLP}$) takes the task query embedding $\mathbf{Q}$ (acquired by pre-trained text encoder) as input and outputs the fusion weight $\mathbf{\alpha} = [\alpha_{\text{dense}}, \alpha_{\text{sparse}}]$ via a softmax function, dynamically adjusting the contribution of each view.  This allows the final topology to adapt to the semantic demands of different tasks, favoring dense propagation when the context is reliable, and leaning toward sparse filtering under uncertainty. Then, we can acquire $\mathbf{M}_{\text{final}}$ by:
\begin{equation}
\resizebox{.85\hsize}{!}{$
\mathbf{M}_{\text{final}} = \alpha_{\text{dense}} \cdot \mathbf{M}_{\text{dense}} + \alpha_{\text{sparse}} \cdot \mathbf{M}_{\text{sparse}}.$}
\end{equation}

The coefficient $\mathbf{M}_{\text{final}}$ is used to sample a discrete communication topology $\mathcal{G}$ that determines how messages flow between agents during inference, following the Bernoulli edge sampling strategy proposed in~\citet{zhang2024cut}.

\paragraph{Model Optimization}
To optimize this structure generation process, we aim to maximize the utility of the resulting communication topology, which measures the correctness of the final output $\mathcal{O} = \mathcal{G}(\mathcal{Q}) $ under the learned communication topology. However, the utility function $\phi(\cdot)$ is typically non-differentiable under our task setting since it depends on downstream black-box reasoning (e.g., querying LLMs). Hence, we adopt a standard policy gradient method to estimate gradients and update model parameters:
\begin{equation}
       \resizebox{.85\hsize}{!}{$\nabla_{\Theta}\mathbf{E}_{\Theta\sim\Omega}\Big[\phi\big(\mathcal{G}(\mathcal{Q})\big)\Big]\approx \frac{1}{M}\sum_{m=1}^{M}\phi(\mathcal{O}_{m})\nabla_{\Theta}\log(P(\mathcal{G}_{m}))$},
   \end{equation}
where the $\Theta\sim\Omega$ are learnable parameters in \ourmethod, $P(\mathcal{G}_{m})$ calculates the probability of $\mathcal{G}_{m}$ being sampled. $\phi(\mathcal{O}_{m})$ can be treated as the task-specific reward of the final answer produced under current topology. The pseudo-code algorithm of \ourmethod is provided in Appendix~\ref{appe:method_detail}.

\begin{table*}[t]
\vspace{-3mm}
\centering
\caption{Performance comparison (\%) on six benchmarks. The best results are highlighted in \textbf{bold}.}\vspace{-2mm}
\label{tab:performance}
\renewcommand{\arraystretch}{0.8} 
\resizebox{\textwidth}{!}{%
\begin{tabular}{l|cccccc|c}
\toprule
\textbf{Method}       & \textbf{MMLU} & \textbf{GSM8K} & \textbf{AQuA} & \textbf{MultiArith} & \textbf{SVAMP} & \textbf{HumanEval} & \textbf{Avg.} \\
\midrule
Vanilla                                & 80.39      & 82.30          & 71.06              & 93.09        & 86.55          & 71.39             & 80.80         \\ \midrule
CoT                              & 81.69 \scriptsize$\uparrow$1.30 &
86.50 \scriptsize$\uparrow$4.20 &
73.58 \scriptsize$\uparrow$2.52 &
93.25 \scriptsize$\uparrow$0.16 &
87.36 \scriptsize$\uparrow$0.81 &
74.67 \scriptsize$\uparrow$3.28 &
82.84 \scriptsize$\uparrow$2.04          \\
SC (CoT)               &
83.66 \scriptsize$\uparrow$3.27 &
81.60 \scriptsize$\downarrow$0.70 &
75.63 \scriptsize$\uparrow$4.57 &
94.12 \scriptsize$\uparrow$1.03 &
88.59 \scriptsize$\uparrow$2.04 &
79.83 \scriptsize$\uparrow$8.44 &
83.91 \scriptsize$\uparrow$3.11          \\
 \midrule
 Chain                              &83.01 \scriptsize$\uparrow$2.62 &
88.30 \scriptsize$\uparrow$6.00 &
74.05 \scriptsize$\uparrow$2.99 &
93.27 \scriptsize$\uparrow$0.18 &
87.17 \scriptsize$\uparrow$0.62 &
81.37 \scriptsize$\uparrow$9.98 &
84.53 \scriptsize$\uparrow$3.73          \\
Tree              &
81.04 \scriptsize$\uparrow$0.65 &
85.20 \scriptsize$\uparrow$2.90 &
71.23 \scriptsize$\uparrow$0.17 &
93.68 \scriptsize$\uparrow$0.59 &
88.91 \scriptsize$\uparrow$2.36 &
80.53 \scriptsize$\uparrow$9.14 &
83.43 \scriptsize$\uparrow$2.63        \\
Complete                           &
82.35 \scriptsize$\uparrow$1.96 &
80.10 \scriptsize$\downarrow$2.20 &
72.95 \scriptsize$\uparrow$1.89 &
94.53 \scriptsize$\uparrow$1.44 &
84.01 \scriptsize$\downarrow$2.54 &
79.03 \scriptsize$\uparrow$7.64 &
82.16 \scriptsize$\uparrow$1.36         \\
Random               &
84.31 \scriptsize$\uparrow$3.92 &
86.90 \scriptsize$\uparrow$4.60 &
76.48 \scriptsize$\uparrow$5.42 &
94.08 \scriptsize$\uparrow$0.99 &
87.54 \scriptsize$\uparrow$0.99 &
82.66 \scriptsize$\uparrow$11.27 &
85.33 \scriptsize$\uparrow$4.53        \\
LLM-Debate                           &
84.96 \scriptsize$\uparrow$4.57 &
91.40 \scriptsize$\uparrow$9.10 &
77.65 \scriptsize$\uparrow$6.59 &
96.36 \scriptsize$\uparrow$3.27 &
90.11 \scriptsize$\uparrow$3.56 &
84.70 \scriptsize$\uparrow$13.31 &
87.53 \scriptsize$\uparrow$6.73       \\
 \midrule
AgentPrune                 & 85.07 \scriptsize$\uparrow$4.57      & 91.10 \scriptsize$\uparrow$8.80          & 80.51 \scriptsize$\uparrow$9.45       & 94.65 \scriptsize$\uparrow$1.56               & 90.58 \scriptsize$\uparrow$4.03          & 86.75 \scriptsize$\uparrow$15.36              & 88.09 \scriptsize$\uparrow$7.29         \\
AgentDropout              & 85.62 \scriptsize$\uparrow$5.23        & 91.70 \scriptsize$\uparrow$9.40          & 80.94 \scriptsize$\uparrow$9.88        & 95.60 \scriptsize$\uparrow$2.51               & 91.04 \scriptsize$\uparrow$4.49         & 85.98 \scriptsize$\uparrow$14.59              & 88.48 \scriptsize$\uparrow$7.68         \\
G-designer             & 86.92 \scriptsize$\uparrow$6.53       & 93.80 \scriptsize$\uparrow$11.50         & 81.60 \scriptsize$\uparrow$10.54         & 96.50 \scriptsize$\uparrow$3.41            & 93.10 \scriptsize$\uparrow$6.55          & 88.33 \scriptsize$\uparrow$16.94             & 90.04 \scriptsize$\uparrow$9.24        \\  \midrule
\ourmethod             & \textbf{88.90 \scriptsize$\uparrow$8.51}       & \textbf{95.20 \scriptsize$\uparrow$12.90}          & \textbf{83.49 \scriptsize$\uparrow$12.43}         & \textbf{96.83 \scriptsize$\uparrow$3.74}               & \textbf{94.70 \scriptsize$\uparrow$8.15}         & \textbf{89.15 \scriptsize$\uparrow$17.76}              & \textbf{91.38 \scriptsize$\uparrow$10.58}          \\  
\bottomrule
\end{tabular}
}
\vskip -1 em
\end{table*}

\section{Experiments}
\subsection{Experimental Setup}
\paragraph{Datasets} We evaluate \ourmethod on six benchmarks from three domains: general reasoning: MMLU~\cite{hendrycks2020measuring}, mathematical reasoning: GSM8K~\cite{cobbe2021training}, MultiArith~\cite{roy2016solving}, SVAMP~\cite{patel2021nlp}, and AQuA~\cite{ling2017program}, and code generation: HumanEval~\cite{chen2021evaluating}, following the evaluation setup adopted in G-designer~\cite{zhang2024g}. Details are in the Appendix~\ref{app:detail}.

\noindent\textbf{Baselines} \ \  We consider a broad range of baselines covering both single-agent prompting strategies and multi-agent communication: 1) single-agent methods such as CoT~\cite{wei2022chain}, Self-Consistency~\cite{wang2022self}; 2) multi-agent systems with fixed topologies including Chain, Tree, Complete Graph, and Random Graph~\cite{qian2024scaling}; 3) LLM-Debate~\cite{du2023improving}; and 4) automated design approaches including AgentPrune~\cite{zhang2024cut}, AgentDrop~\cite{wang2025agentdropout}, and G-Designer~\cite{zhang2024g}. 

\noindent \textbf{Implementation Details} \ \ We use GPT-4o throughout all experiments, accessed via the OpenAI API. A summarizer agent is designated to aggregate the dialogue history and produce the final answer, where the round $K=3$. This also corresponds to the number of GNN layers used. For agent representation, we implement the $\operatorname{NodeEncoder}(\cdot)$ using the pretrained \texttt{all-MiniLM-L6-v2}, with an embedding dimension $D=384$. We use 6 agents for the complex reasoning benchmark MMLU, 5 agents for HumanEval, and 4 agents for all mathematical reasoning benchmarks. Each experiment is optimized using $B \in \{40,60\}$ query samples.
\subsection{Performance Comparison}
\textbf{\ourmethod achieves state-of-the-art performance on all six benchmarks}. \ \ As shown in Table \ref{tab:performance}, \ourmethod obtains the highest average accuracy of \textbf{91.38\%}, outperforming all single-agent and multi-agent baselines which adopt the predefined or LLM-generated topology. Specifically, predefined topologies achieve an average performance of at most \textbf{85.53\%} (e.g., LLM-Debate), while automated design methods perform better, with G-designer reaching up to \textbf{90.04\%}. Even against recent best topology optimization methods, \ourmethod still achieves an average performance improvement of $\textbf{1.34\%}$. Experimental results verify that \ourmethod generates a better topology and achieves accurate and reliable decisions by balancing insight and error propagation. We provide the case study in the Appendix~\ref{app:case_study}.

\begin{table}[t]
\centering
\caption{Results of ablation study.}

\label{tab:ablation}
\renewcommand{\arraystretch}{0.8} 
\vspace{-3mm}
\resizebox{\columnwidth}{!}{%
\begin{tabular}{l|ccc|c}
\toprule
\textbf{Method}       & \textbf{MMLU} & \textbf{GSM8K} & \textbf{HumanEval} & \textbf{Average} \\ 
\midrule
Vanilla  & 80.39 & 82.30 & 71.39 & 78.02 \\
\ourmethod  & \textbf{88.90} & \textbf{95.20} & \textbf{89.15} & \textbf{91.08} \\ \midrule
w/o Dense & 84.85 & 91.10 & 86.57 & 87.51 \\
w/o Sparse & 86.17 & 93.50 & 88.26 & 89.31 \\
w/o Fusion & 85.23 & 92.80 & 87.07 & 88.37 \\

\bottomrule

\end{tabular}
}
\vskip -1 em
\end{table}
\subsection{Ablation Study}
\textbf{All modules in \ourmethod are critical to maintaining strong reasoning performance.} We conduct ablation studies on MMLU, GSM8K, and HumanEval to evaluate the impact of different communication components: (1) \textbf{w/o Dense}, which removes the dense GNN branch; (2) \textbf{w/o Sparse}, which excludes the sparse GNN branch; and (3) \textbf{w/o Fusion}, which replaces the fusion module with a direct addition operation. As shown in Table~\ref{tab:ablation}, removing the dense GNN branch leads to the largest performance drop. For example, on GSM8K, accuracy decreases by $\textbf{4.1\%}$. The absence of either the sparse GNN branch or the fusion component also results in comparable degradation. These results suggest that the three modules interact synergistically, and removing any of them weakens the overall reasoning capacity of MAS. 
\vspace{-5pt}
\subsection{Token Efficiency}
\textbf{\ourmethod achieves lower token cost with higher performance.}
As shown in Figure~\ref{subfig:tokenMMLU} and ~\ref{subfig:tokenGSM8K}, although our approach does not explicitly pursue sparsity, it achieves comparable cost to sparsity-based G-designer while delivering significantly higher accuracy. This observation holds across both MMLU and GSM8K, where \ourmethod maintains efficient communication without compromising reasoning quality. Compared to predefined communication topologies such as fully connected graphs and LLM-Debate, \ourmethod substantially reduces token consumption which highlighting its ability to eliminate redundant interactions while preserving critical information flow.
\vspace{-5pt}
\subsection{Robustness Analysis}
\begin{figure}[t]
\vspace{-3mm}
  \centering
  \subfloat[Token cost of MMLU]{%
    \label{subfig:tokenMMLU}%
    \includegraphics[width=0.5\columnwidth]{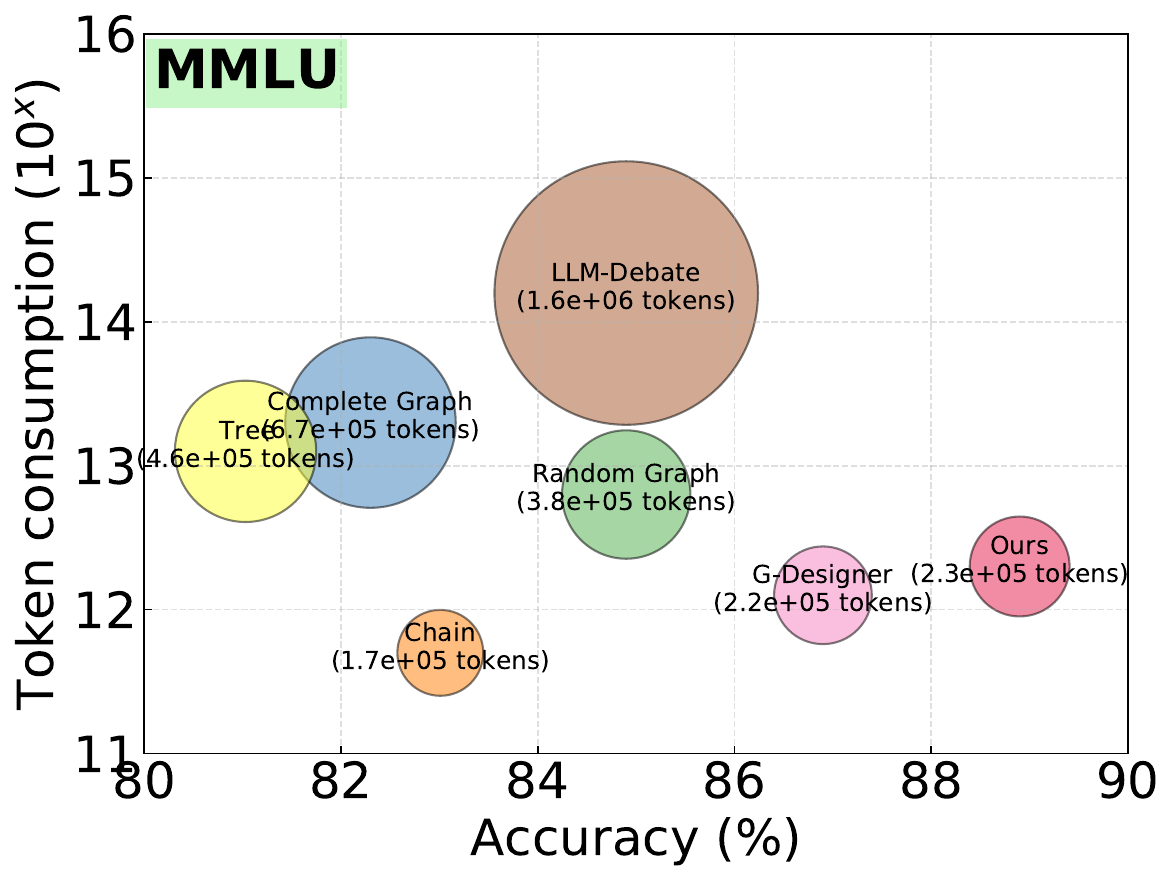}%
  }\hfill
  \subfloat[Token cost of GSM8K]{%
    \label{subfig:tokenGSM8K}%
    \includegraphics[width=0.5\columnwidth]{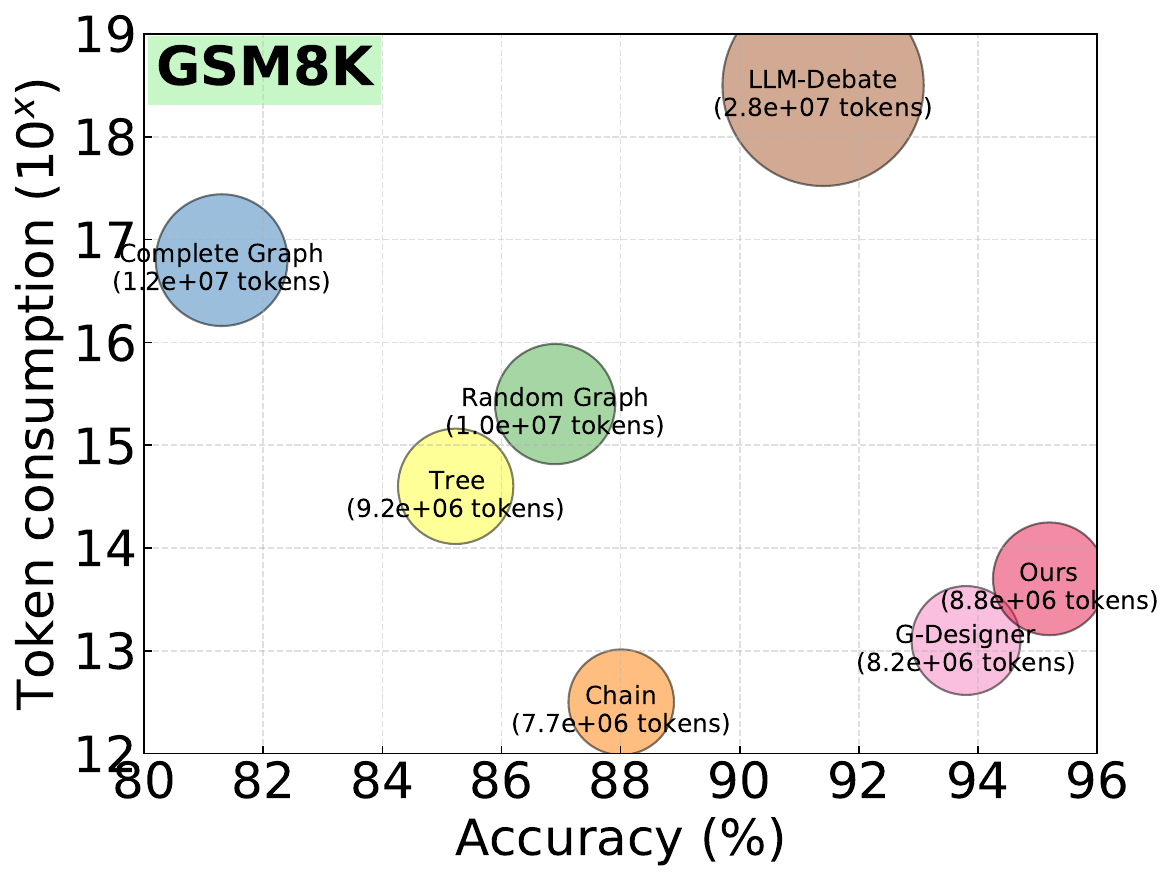}%
  }

  \vspace{-1em} 

  \makebox[\columnwidth][s]{%
    \subfloat[Robustness under attack]{%
      \label{subfig:attack}%
      \includegraphics[width=\columnwidth]{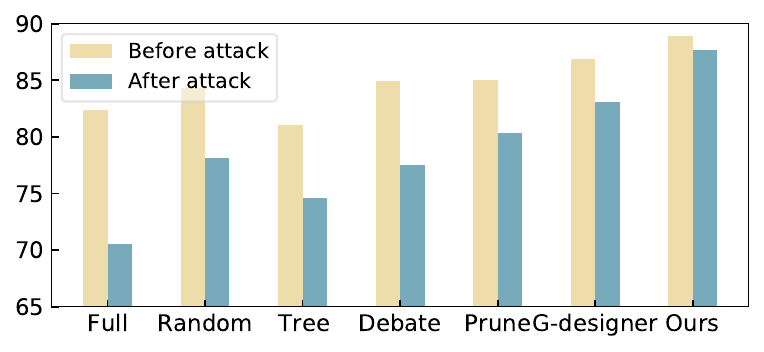}%
    }%
  }
\vspace{-3mm}
  \caption{Experiments on token consumption and robustness against prompt injection attacks.}
  \label{fig:three_plots_exp}
\vspace{-3mm}
\end{figure}

\textbf{\ourmethod demonstrates strong robustness against adversarial prompt attacks.}
Following~\citet{zhuge2024gptswarm}, we simulate a system prompt attack by injecting adversarial prompts into a single agent. As shown in Figure~\ref{subfig:attack}, simple topologies such as chain and tree graphs suffer substantial degradation under such attacks, with performance drops up to $\textbf{11.8\%}$.  In contrast, \ourmethod exhibits strong robustness: the accuracy only drops by the least decrease of $\textbf{1.24\%}$. This robustness arises from our balanced topology design, which mitigates error propagation through bottleneck structures while promoting rapid consensus via densely connected reliable agents. 

\section{Conclusion}
\vspace{-7pt}
In this paper, we present \ourmethod, a topology optimization framework for LLM-based multi-agent systems, grounded in causal analysis of information propagation. By identifying that moderately sparse structures best balance error suppression and insight preservation, \ourmethod adaptively fuses dense and sparse views to form task-specific topologies. Experiments across reasoning, math, and code tasks show that \ourmethod consistently improves performance, enhances robustness, and reduces communication overhead.
\section*{Limitation}
While our method demonstrates strong performance on reasoning, math, and code generation tasks, the scope of evaluation remains limited. To better assess the generalizability of our framework, future work should explore more diverse task domains, such as real-world decision-making and open-domain dialogue. In addition, the current design relies on a fixed set of predefined agent roles and manually crafted prompts, which may limit adaptability in unfamiliar or evolving scenarios. Automatically discovering agent roles and optimizing prompts is a promising direction for improving robustness and flexibility.
\section*{Ethics Statement}
Our research focuses exclusively on scientific questions, with no involvement of human subjects, animals, or environmentally sensitive materials. Therefore, we foresee no ethical risks or conflicts of interest. We are committed to maintaining the highest standards of scientific integrity and ethics to ensure the validity and reliability of our findings.
\bibliography{acl_latex}

\newpage
\clearpage
\appendix

\section{Related Work}
\subsection{LLM-based Multi-Agent Systems}
Large Language Models (LLMs) have demonstrated remarkable capabilities across a wide range of tasks, from reasoning~\cite{wang2024unleashing,wang2022self} and planning to dialogue~\cite{hu2024agentgen,yi2024survey} and programming~\cite{ishibashi2024self,zhang2024codeagent}. Extending from single-agent settings, recent work has explored organizing multiple LLM-based agents into collaborative systems, unlocking new potential through inter-agent interaction~\cite{guo2024large}. Representative systems adopt diverse coordination strategies, including sequential reasoning pipelines~\cite{wei2022chain}, debate-based role playing~\cite{li2024improving}, and centralized planning~\cite{zhuge2024gptswarm}. These advances highlight the importance of inter-agent communication in scaling LLM capabilities to more complex, open-ended tasks.
\subsection{MAS as Graphs}
The effectiveness of LLM-based MAS has been closely tied to the quality of their communication topologies~\cite{zhuge2024gptswarm}. Graphs provide a natural and expressive abstraction for modeling inter-agent interactions, enabling both structured information flow and flexible coordination~\cite{hu2024learning,zhou2025multi}. Early approaches adopt fixed topologies such as chains, stars, or fully connected graphs~\cite{qian2024scaling}, each offering different trade-offs in information sharing and control~\cite{li2024improving}. More recent research explores learnable communication graphs to enhance execution efficiency and scalability. For example, AgentPrune~\cite{zhang2024cut} introduces task-specific sparse masks to suppress redundant communication; AgentDrop~\cite{wang2025agentdropout} applies stochastic dropout to nodes and edges to improve robustness; and G-Designer~\cite{zhang2024g} dynamically generates query-dependent topologies for better task adaptation. While these methods improve computational efficiency and economic viability by learning compact and adaptive communication structures, they generally treat sparsity as a cost-saving mechanism without explicitly analyzing when or why sparse or dense topologies lead to better multi-agent outcomes.
\section{Pseudo-code for \ourmethod}\label{appe:method_detail}
Algorithm~\ref{alg:gdesigner} provides the detailed pseudo-code of \ourmethod, which takes the current query as input. \ourmethod comprises a lightweight language encoder, two GNNs, and a gated fusion network, ensuring efficiency during execution. In training, a subset of queries is used to construct communication topologies via \ourmethod, followed by multi-agent decision-making based on the generated topologies.
\begin{table}[h]
\centering
\caption{Comparison of accuracy, time, and cost across different agent configurations. We sample 1000 questions from MMLU benchmark
and utilize GPT-3.5 as the base LLM.}
\label{tabel:cost}
\renewcommand{\arraystretch}{1.1}  
\begin{tabular}{l|ccc}
\toprule
\#Agents            & \textbf{5}      & \textbf{7}     & \textbf{9}     \\
\midrule
\textbf{Chain}     &               &               &               \\                                       
Accuracy (\%)       & 63.33           & 64.55           & 65.07           \\
Time (min)          & 53.11           & 90.94           & 126.15           \\
Cost (USD)          & 4.18          & 10.04          & 15.38          \\
\midrule
\textbf{Complete Graph} &          &               &               \\
Accuracy (\%)       & 61.83           & 62.16           & 62.71           \\
Time (min)          & 80.85           & 114.29           & 236.44           \\
Cost (USD)          & 5.78          & 14.49          & 31.67          \\
\midrule
\textbf{G-Designer}  &               &               &               \\
Accuracy (\%)       & 64.08           & 65.82           & 66.01           \\
Time (min)          & 57.14           & 94.32          & 138.18          \\
Cost (USD)          & 4.30          & 11.37         & 16.35         \\
\midrule
\textbf{\ourmethod}&               &               &               \\
Accuracy (\%)       & \textbf{65.47}  & \textbf{67.14}  & \textbf{67.82}  \\
Time (min)          & 59.38           & 96.21           & 141.89           \\
Cost (USD)          & 4.36          & 11.97         & 17.02        \\
\bottomrule
\end{tabular}
\end{table}
\section{Scalability in the number of agents}\label{app:agentnum}
To evaluate the scalability of \ourmethod to larger multi-agent settings, we report its performance across systems with 5 to 9 agents, as shown in Table \ref{tabel:cost}. Notably, \ourmethod achieves the most significant performance gains as the number of agents increases. More importantly, while maintaining comparable computational cost and runtime to G-Designer, \ourmethod delivers consistently better results. Specifically, it achieves a $5.11\%$ improvement in accuracy while consuming only about half the token cost of a fully connected topology. These results demonstrate the scalability and potential of our method for enabling large-scale, autonomous multi-agent collaboration.
\begin{figure*}[ht!]    
					\centering    
	\includegraphics[scale=0.45]{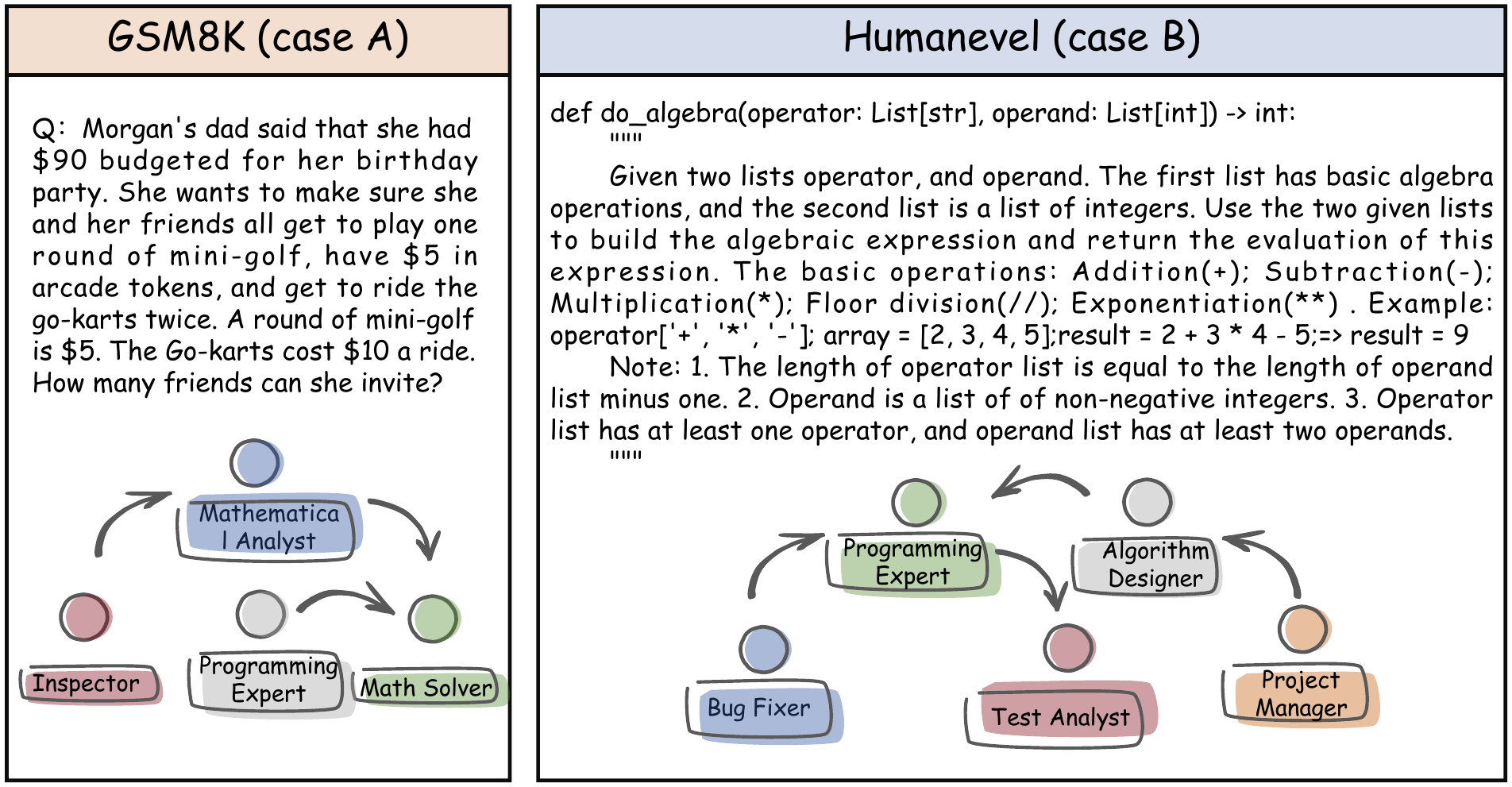}
			
					\caption{Case study of the communication topologies designed by \ourmethod on HumanEval and GSM8K benchmarks.}
					\label{fig:case}           
                    \vskip -1 em
				\end{figure*}

\begin{table*}[ht]
\centering
\caption{Dataset descriptions and statistics.}
\label{tab:dataset_stats}
\begin{tabular}{llllrl}
\toprule
\textbf{Category} & \textbf{Dataset} & \textbf{Answer Type} & \textbf{Metric} & \textbf{\#Test} & \textbf{License} \\
\midrule
General reasoning & MMLU & Multi-choice & Acc. & 153 & MIT License \\ \midrule
\multirow{4}{*}{Math reasoning} & GSM8K & Number & Acc. & 1,319 & MIT License \\ 
 & MultiArith & Number & Acc. & 600 & Unspecified \\
 & SVAMP & Number & Acc. & 1,000 & MIT License \\
 & AQuA & Multi-choice & Acc. & 254 & Apache-2.0 \\ \midrule
Code generation & HumanEval & Code & Pass@1 & 164 & MIT License \\
\bottomrule
\end{tabular}
\end{table*}
\section{Dataset Statistic}\label{app:detail}
We present the dataset statistics in Table \ref{tab:dataset_stats}, following the same experimental setup as G-Designer~\cite{zhang2024g}.
\section{Case Study}\label{app:case_study}
For the HumanEval task (case A in Figure~\ref{fig:case}), which requires symbolic reasoning and program synthesis, \ourmethod constructs a hierarchical communication structure: the \textbf{Project Manager} initiates task planning and forwards it to the \textbf{Algorithm Designer}, who devises logic that is implemented by the \textbf{Programming Expert}. Outputs are then verified by the \textbf{Test Analyst} and corrected by the \textbf{Bug Fixer}, forming a feedback loop that suppresses local errors while ensuring functional accuracy.

For the GSM8K problem (case B in Figure~\ref{fig:case}), calculating how many friends can attend a party under budget constraints. \ourmethod topology to balance efficiency and error control. The \textbf{Mathematical Analyst} models the problem and communicates with the \textbf{Math Solver} for final computation. Simultaneously, the \textbf{Programming Expert} provides structural parsing to the \textbf{Inspector}, who ensures the calculation logic is sound, avoiding propagation of misunderstandings.

These topologies align with our dual-view design by enabling efficient propagation of correct reasoning while suppressing misleading signals through targeted verification pathways.

\begin{algorithm*}[t]
\caption{Designing workflow of \ourmethod}
\label{alg:gdesigner}

\KwIn{Input query $\mathcal{Q}$, a $\mathrm{NodeEncoder}$, two \text{GNN}, a gating function, learning rate $\alpha$}

\For{query $d\in\{1,2,\dots,D'\}$}{\Comment{Establish multi-agent network}
    \For{node $i\in\{1,2,\dots,N\}$}{%
        \(|\) $\mathbf{x}_i \gets \mathrm{NodeEncoder}\bigl(\mathcal{T}(\text{Role}_i), \mathcal{Q}_d\bigr)$\;
    }
    Obtain agent embeddings 
      $\mathbf{X}\gets[\mathbf{x}_1,\mathbf{x}_2,\dots,\mathbf{x}_N]^\top$\;

      Obtain query embeddings 
     \(|\) $\mathbf{Q}_d \gets \mathrm{NodeEncoder}\bigl(\mathcal{Q}_d\bigr)$\;
      \Comment{Forward process of \ourmethod}

    Utilize the fully connected graph and chain to obtain two adjacency matrices $\mathbf{A}_{\text{dense}}$ and $\mathbf{A}_{\text{sparse}}$;
    
    Utilize $\mathbf{A}_{\text{dense}}$ and $\mathbf{A}_{\text{sparse}}$ to 
    encode $\mathcal{G}$ into latent representations and decoding results in two soft masks $\mathbf{M}_{\text{dense}}$ and  $\mathbf{M}_{\text{sparse}}$:
    
    $\mathbf{Z}_{\text{dense}} = \operatorname{GNN}(\mathbf{A}_{\text{dense}},\mathbf{X}),$
    $\mathbf{M}_{\text{dense}} =  \sigma(\mathbf{Z}_{\text{dense}}^{T}\mathbf{Z}_{\text{dense}})$

    $\mathbf{Z}_{\text{sparse}} = \operatorname{GNN}(\mathbf{A}_{\text{sparse}},\mathbf{X}),$
    $\mathbf{M}_{\text{sparse}} =  \sigma(\mathbf{Z}_{\text{sparse}}^{T}\mathbf{Z}_{\text{sparse}})$

    Utilize a gating function to calculate the weight $\alpha$ based on the query $\mathcal{Q}$:

    $\alpha = \text{softmax}(\mathbf{W}_2 \cdot \text{ReLU}(\mathbf{W}_1 \mathbf{Q}_d+ \mathbf{b}_1)) $

    Fusing two masks based on their weights:

    $\mathbf{M}_{\text{final}} = \alpha_{\text{dense}} \cdot \mathbf{M}_{\text{dense}} + \alpha_{\text{sparse}} \cdot \mathbf{M}_{\text{sparse}}.$

    Obtain the new communication topology $\mathcal{G}$ by sampling from $\mathbf{M}_{\text{final}}$

    \Comment{Guide multi-agent collaboration}
    Obtain the the execution order of agents by topological sorting $\sigma = [v_{\sigma_1}, \dots, v_{\sigma_N}]$
    
    \For{$t\leftarrow 1$ \KwTo\ $K$}{%
        \For{node $i\in \sigma $}{%
        $\mathcal{R}_{i}^{(t)} = v_{i}(\mathcal{P}_{\text{sys}},\mathcal{P}^{(t)}_{\text{usr}})$\;
    $\mathcal{P}^{(t)}_{\text{usr}}=\{\mathcal{Q}\} \cup \{\mathcal{R}_{j}^{(t)} | v_j\in \mathcal{N}(v_{i})\} $\;
        }
        $\mathcal{O}^{(t)} = \operatorname{Aggregate}(\mathcal{R}_{i}^{(t)} \dots \mathcal{R}_{n}^{(t)})$\;
    }

    \Comment{Update G-Designer parameters}
    $\Theta^{d+1}\gets \Theta^d - \alpha\,\nabla_{\Theta}\mathcal{L}_{\mathrm{\ourmethod}}$\;
}
\end{algorithm*}
\end{document}